\documentclass[sigconf]{acmart}
\pdfoutput=1
\usepackage{makecell}
\usepackage{booktabs}
\usepackage{graphics}
\usepackage{xspace}
\usepackage{subfig}
\usepackage{wasysym}
\usepackage{graphicx}
\usepackage{amsfonts}
\usepackage{xcolor}
\usepackage{CJKutf8}
\usepackage{multirow}
\usepackage{multicol}
\usepackage{array}
\usepackage{xspace}
\usepackage{subfig}
\usepackage{amsmath}
\usepackage{float}
\usepackage{hyphenat}
\usepackage{pgfplots}
\usepackage{tikz}
\usepackage[utf8]{inputenc}
\usepackage{newunicodechar}
\usepackage{graphicx,stackengine,scalerel}

\restylefloat{table}

\def\BibTeX{{\rm B\kern-.05em{\sc i\kern-.025em b}\kern-.08emT\kern-.1667em\lower.7ex\hbox{E}\kern-.125emX}}

\copyrightyear{2018}
\acmYear{2018}
\setcopyright{acmlicensed}
\acmConference[Woodstock '18]{Woodstock '18: ACM Symposium on Neural Gaze Detection}{June 03--05, 2018}{Woodstock, NY}
\acmPrice{15.00}
\acmDOI{10.1145/1122445.1122456}
\acmISBN{978-1-4503-9999-9/18/06}

\begin{document}

\author{Chang-You Tai}
\affiliation{%
   \institution{Academia Sinica}
   \city{Taipei}
   \country{Taiwan}}
\email{johnnyjana730@gmail.com}

\author{Meng-Ru Wu}
\affiliation{%
   \institution{Academia Sinica}
   \city{Taipei}
   \country{Taiwan}}
\email{ray7102ray7102@gmail.com}

\author{Yun-Wei Chu}
\affiliation{%
   \institution{Academia Sinica}
   \city{Taipei}
   \country{Taiwan}}
\email{yunweichu@gmail.com}

\author{Shao-Yu Chu}
\affiliation{%
   \institution{Academia Sinica}
   \city{Taipei}
   \country{Taiwan}}
\email{shaoyu0966@gmail.com}

\author{Lun-Wei Ku}
\affiliation{%
   \institution{Academia Sinica}
   \city{Taipei}
   \country{Taiwan}}
\email{lwku@iis.sinica.edu.tw}

\title{GraphSW: a training protocol based on stage-wise training \\ for GNN-based Recommender Model}

\begin{abstract}
Recently, researchers utilize Knowledge Graph (KG) as side information in recommendation system to address cold start and sparsity issue and improve the recommendation performance. Existing KG-aware recommendation model use the feature of neighboring entities and structural information to update the embedding of currently located entity. Although the fruitful information is beneficial to the following task, the cost of exploring the entire graph is massive and impractical. In order to reduce the computational cost and maintain the pattern of extracting features, KG-aware recommendation model usually utilize fixed-size and random set of neighbors rather than complete information in KG. Nonetheless, there are two critical issues in these approaches: First of all, fixed-size and randomly selected neighbors restrict the view of graph. In addition, as the order of graph feature increases, the growth of parameter dimensionality of the model may lead the training process hard to converge. To solve the aforementioned limitations, we propose GraphSW, a strategy based on stage-wise training framework which would only access to a subset of the entities in KG in every stage. During the following stages, the learned embedding from previous stages is provided to the network in the next stage and the model can learn the information gradually from the KG. We apply stage-wise training on two SOTA recommendation models, RippleNet and Knowledge Graph Convolutional Networks (KGCN). Moreover, we evaluate the performance on six real world datasets, Last.FM 2011, Book-Crossing,movie, LFM-1b 2015, Amazon-book and Yelp 2018. The result of our experiments shows that proposed strategy can help both models to collect more information from the KG and improve the performance. Furthermore, it is observed that GraphSW can assist KGCN to converge effectively in high-order graph feature.

\end{abstract}

\maketitle

\section{Introduction}

To address the matter of the cold-start problem and the sparsity of user-item interactions in CF-based recommendation model, many researchers take the Knowledge Graph (KG) as side information. Because KG, which introduces semantic relatedness among items, contains fruit information and connections between items, it can enhance the performance of recommendation system\cite{Hu:2018:LMB:3219819.3219965,Huang:2018:ISR:3209978.3210017,Sun:2018:RKG:3240323.3240361,Zhang:2016:CKB:2939672.2939673,Zhao:2017:MBR:3097983.3098063}. Recent KG-aware recommendation systems can be roughly classified into three categories: embedding-based methods\cite{DBLP:journals/corr/abs-1902-06236_emb,Huang:2018:ISR:3209978.3210017_emb,Zhang:2016:CKB:2939672.2939673_emb,Wang:2018:DDK:3178876.3186175_emb}, path-based methods\cite{Hu:2018:LMB:3219819.3219965_path,Sun:2018:RKG:3240323.3240361_path,DBLP:journals/corr/abs-1811-04540_path,Yu:2014:PER:2556195.2556259_path,Zhao:2017:MBR:3097983.3098063_path} and Graph Neural Network (GNN) based methods\cite{DBLP:journals/corr/abs-1803-03467,DBLP:journals/corr/abs-1904-12575,DBLP:journals/corr/abs-1905-04413,DBLP:journals/corr/abs-1905-07854,DBLP:journals/corr/abs-1904-10322,DBLP:journals/corr/abs-1905-08108}. Because GNN-based recommendation systems utilize GNN architecture and can realize end-to-end training to exploit high-order information of KG, GNN-based methods eliminate the limitation of embedding-based methods and path-based method. Many researchers study GNN-based recommendation model: Rex et al. \cite{DBLP:journals/corr/abs-1806-01973} applies GNNs on bipartite graph and build recommendation model which is successively deployed at Pinterest. Wu et al. \cite{kdir18GraphConvolutionalMatrix} utilizes multihop neural network structure transform the signals into user/item representations. Wang et al. propose RippleNet\cite{DBLP:journals/corr/abs-1803-03467} and KGCN\cite{DBLP:journals/corr/abs-1904-12575}. RippleNet is a memory-network-like model which propagates items within paths rooted at each users’ potential preferences to produce user representations. KGCN utilize neighborhood aggregation to calculate the item representation. In addition, neighborhood aggregation can be extended to multiple hops away and allow model to capture high-order and long-distance entity dependencies. Wang et al. propose Knowledge Graph Attention Network (KGAT) \cite{DBLP:journals/corr/abs-1905-07854}, which utilize attention network on KG and exploit the user-item graph structure by recursively propagating embeddings.

However, Graph Convolutional Network has neighbor explosion issue when GCN aggregates the neighborhoods nodes. In GNN-based recommendation model, each node's representation in the current layer is aggregated from its neighbors' representation previous layer. As the hop number increase, the multi-hop neighbors would cost huge computation resource. To solve that issue, current GNN-based recommendation, such as PinSage\cite{DBLP:journals/corr/abs-1806-01973}, RippleNet\cite{DBLP:journals/corr/abs-1803-03467} and KGCN\cite{DBLP:journals/corr/abs-1904-12575}, would adopt "fixed-size" strategy that in each layer model would only sample a fixed-size set of neighbors, instead of using full neighborhood sets, in previous layer to reduce computation resource. To use more neighborhoods information, in each minibatch iteration, PinSage would resample another fixed-size set of neighbor for each layer. However, in original paper, it don't discuss the performance gain from that resampling strategy and the but only discusses the trade-off of performance and runtime when the different size of neighborhood is set. Could that strategy be applied to different model and dataset? In addition, the statistic of dataset is not included and we don't know the exact number of entity-relation-entity triplets information are used when model achieves best performance.

In addition to neighbor explosion issue, GNN-based recommendation model, such as KGCN which has architecture similar to PinSage, would face issue that model is hard to converge as a order of graph feature increases. Because KGCN's architecture is designed to automatically capture both high-order structure and semantic information in KG, massive noise entity and the growth of parameter dimensionality of the KGCN would lead the training process hard to converge when the order of graph feature increases\cite{DBLP:journals/corr/abs-1904-12575}. 
To assist the speed of convergence and preclude the noisy feature in the beginning of deep neural network's training process, the strategy of stage-wise training is proposed\cite{pmlr-v44-Barshan2015}. The learning process of stage-wise training is broken down into a number of related sub-tasks and the training data is presented to the network gradually. In addition, learned feature in previous stage is extracted and transferred to the next stage in order to gradually absorb the sharing knowledge between task and finally achieve better predictions during training process. Stage-wise training has been used in different applications such as multi-task learning\cite{DBLP:journals/corr/abs-1803-08396}, feature extraction layer\cite{pmlr-v44-Barshan2015} and multi-model recognition\cite{DBLP:journals/corr/EitelSSRB15}.

On other hands, we found that in the original training protocol of GNN-based recommendation models, KGCN and RippleNet don't adpot the resampling strategy. As a result, we think these two models are the good choice to study the exact performance change from resampling strategy. In this paper, we aim to conduct the comprehensive study on the exact performance change from resampling strategy on different variant and handle the difficult convergence of KGCN as the order of graph feature increases. As a result, we propose GraphSW, a strategy based on stage-wise training framework. In every stage, KGCN and RippleNet are fed with only a small subset of entity in KG instead of massive entity which may allow the model to easily find crucial information. During the following stages, the learned embedding from previous stages is provided to the network and the model can learn the information from KG gradually. Empirically, we evaluate and train two recommendation models, RippleNet and KGCN, with stage-wise learning on six real-world datasets, i.e., movie, LFM-1b 2015, Amazon-book and Yelp 2018. The result shows that stage-wise learning allows KGCN and RippleNet to collect more information from KG and improves recommendation performance on all datasets. We conduct comprehensive study on the performance gain on different variant and to our surprise, we found that KGCN's performance achieves best when the neighbor sampling size is small and even KGCN don't all information in KG. In addition, for KGCN, we found that stage-wise training can mitigate the difficult convergence issue of model as the order of graph increases. As hop number grows to 4, the average improvement of stage-wise training on AUC is 34.8$\%$, 17.5$\%$, 2.3$\%$, 2.2$\%$, 8.2$\%$ and 0.9$\%$ for Last.FM 2011, Book-Crossing, MovieLens-1M, LFM-1b 2015, Amazon-book and Yelp 2018 respectively.

In summary, this work includes four contributions. 
\begin{itemize}
\item With GraphSW, we conduct comprehensive study on performance gain when more KG information is used on six real-world datasets consisting of different size of KG.

\item In general, we find that GraphSW improve performance of KGCN and RippleNet. However, to our surprise, it is found that KGCN's performance achieves best when small neighbor sampling size is set.

\item Because the GraphSW assist KGCN to collect useful information and preclude noisy one, the difficult convergence issue can be addressed in high order graph feature.

\item We release the code of KGCN and RippleNet with stage-wise learning to researchers for validating the reported results and conducting further research. The code and the data are available at https://github.com/mengruwu/graphsw.

\end{itemize}

\section{GraphSW}
In this section, we introduce the proposed GraphSW, a training protocol based on stage-wise training. First, we briefly describe the problem formulation of KG aware recommendation. Then, we present the design of the proposed stage-wise training on KGCN.

\begin{figure*}[htb]
\centering
\includegraphics[scale=0.42, trim={0 0 0 0}]{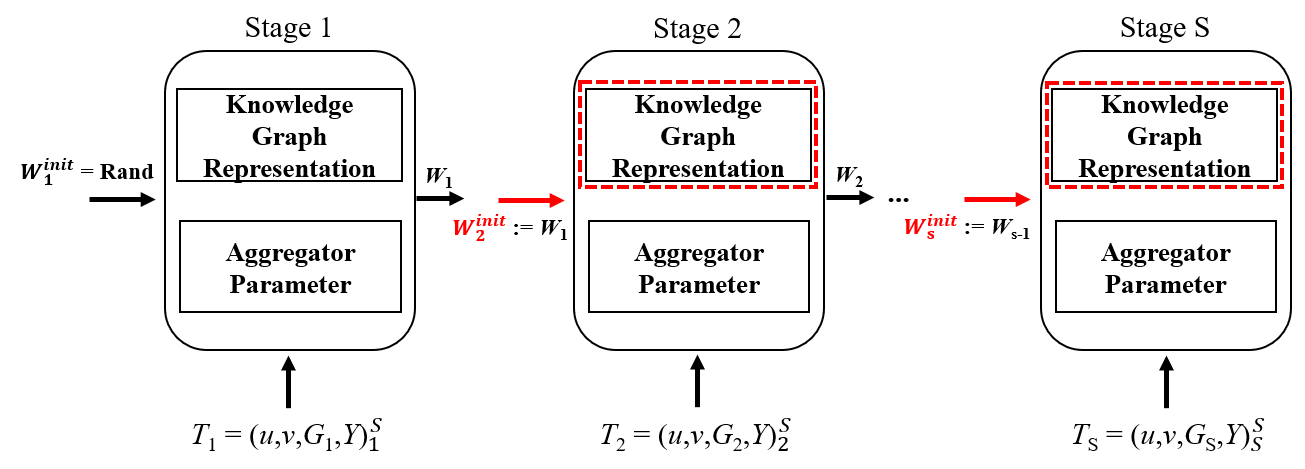}
\caption{Schematic diagram of stage-wise training on KGCN}
\label{fig:stage_wise_plot}
\end{figure*}

\subsection{Recommendation Task Formulation}

In this section, we introduce the task formulation of KG-aware recommendation system. The sets of users and items are denoted as \(\mathcal{U} = \{u_1,u_2...\}\) and \(\mathcal{V} = \{v_1,v_2...\}\) and the user-item interaction matrix \(\mathbf{Y}=\{y_{uv} \mid u \in \mathcal{U}, v \in \mathcal{V}\}\) is defined according to user's implicit feedback. If user $u$ has engaged with item \(v\), \(y_{uv}\) would be recorded as \(y_{uv}=1\); otherwise \(y_{uv}=0\) . In addition, knowledge graph used by KG-aware recommendation model is denoted as $\mathcal{G}$ and is comprised of entity-relation-entity triples \(\{(h,r,t)|h,t \in \mathcal{E}, r \in \mathcal{R}\}\), where $h$, $r$ and $t$ denote head, relation and tail of a knowledge triple and $\mathcal{E}$ and $\mathcal{R}$ denote the set of entities and relations in KG. $\mathcal{N}(v)$ denotes as number neighborhood node of item $v$ and the set of neighborhood node of item $v$ is denoted as $\mathcal{S}(v)\triangleq\{ e \mid e \sim \mathcal{N}(v)\}$ and $|\mathcal{S}(v)|= K$, where $K$ is neighbors sampling size. Recommendation model is used to predict whether user $u$ has interest in item which user has no interaction history before, and the ultimate goal of KG-aware recommendation model is to learn prediction function \(\hat{y}_{uv} = \mathcal{F}(u,v;\Theta, \mathcal{G})\), where $\Theta$ is model parameter and $\hat{y}_{uv}$ is the probability that user $u$ will engage with item $v$.

\subsection{Training Stage}

In stage-wise training, the training set in stage $s$, is denoted as \(T_s = {(u,v,G_{s},Y)}\) , where $s$ is current stage of training and \(s \in \{1,2,...,S\}\), $u \in \mathcal{U}$ and $v \in \mathcal{V}$ are user-item pairs, $Y$ is entire user-item interaction matrix, and $G_{s}$ is fixed-size set of neighbors of all items in knowledge graph at stage $s$ and $G_{s}=\sum_{n=v}^{\mathcal{V}}\mathcal{S}(n)$. We define KG-aware recommendation's learning algorithm as \(A(., .)\) and the learned parameter was defined as $W_{s}$,
\begin{equation} 
   W_{s} = A(T_{s},W^{init}_{s}) \\ 
\end{equation}
where, $W^{init}_{s}$ is the initial value of the parameter, and the first-stage parameter $W^{init}_{1}$ is randomly initialized. The connecting successive stages would be defined as follow:

\begin{equation}
W^{init}_{s+1} := W_{s}, \forall s \in \{1,...,S-1\}
\end{equation}
where, $W_{s}$ is learned parameter of whole model and $W^{init}_{s+1}$ is transferred parameter. In each training stage, we would save $W_{s}$ and transfer part of parameter $W^{init}_{s+1}$ from previous stages to next stages.

\subsection{Training Protocol}
 The whole training protocol is illustrated in Figure~\ref{fig:stage_wise_plot}. The parameter of GNN-based recommendation system can be roughly classified into two parts: knowledge graph representation and aggregator parameter. Considering the limited training performance resulted from high parameter dimensionality, we train the model parameters in different stages. Because original model is designed to increase computation efficiency, only fix-size set of neighborhood in KG, $G_{s}$, is sampling by aggregator, and the model can only collect part of information in KG during each training stage. As a result, we first fine tune the knowledge graph representation and collect more entity information of KG representation to explore the graph more comprehensively. During following training stages, we extract all model parameters $W_{s}$ and transfer only learned knowledge graph representation $W^{init}_{s+1}$ from previous stages to next stage. In addition, the model would randomly sample another set of neighborhood in KG, $G_{s+1}$, to collect more entity information to KG representation in the next stage. After knowledge graph representation is well trained, it would be utilized to fine tune the remaining aggregator part and the prediction would perform better because of the comprehensive view of graph.

\section{Experiments AND RESULTS}
\label{sec:exp}

In this section, we evaluate the recommendation performance of RippleNet and KGCN with stage-wise training. First, we introduce the datasets, two SOTA GNN-based Recommender models, model settings and experiment setup. Then, we present and discuss the recommendation performance.

\begin{table*}[t]
\centering
\begin{tabular}{c|cccccc}  & 
\multicolumn{1}{c}{Last.FM2011}  & 
\multicolumn{1}{c}{Book-Crossing}  & 
\multicolumn{1}{c}{MovieLens-1M}  & 
\multicolumn{1}{c}{LFM-1b2015}  & 
\multicolumn{1}{c}{Amazon-book}  & 
\multicolumn{1}{c}{Yelp2018}\\
\hline \# users & 1872 & 17860 & 6036 & 12134 & 6969 & 15933 \\ \# items & 3846 & 14967 & 2445 & 15471 & 9854 & 13873 \\ \# interactions & 42346 & 139746 & 753772 & 2979267 & 552706& 1159936 \\ Avg \# of user click & 22.6 & 7.8 & 124.9 & 152.3 & 79.3 & 72.8\\ Avg \# of click item  & 11.0 & 9.3 & 308.3 & 119.4 & 56.1& 83.6 \\ \hline KG \#entities & 9366 & 77903 & 182011 & 106389 & 113487 & 136499 \\ KG \#relations & 60 & 25 & 12 & 9 & 39 & 42\\ KG \#triples & 15518 & 151500 & 1241995 & 464567 & 2557746 & 1853704 \\
\end{tabular}
\caption{dataset basic statistic}
\label{tb:datasetstatistic}
\vspace{-1em}
\end{table*}

\subsection{Datasets}

To evaluate proposed strategy, we utilize six real-world datasets: Last.FM 2011, Book-Crossing, MovieLens-1M, LFM-1b 2015, Amazon-book and Yelp 2018 which are publicly available\cite{DBLP:journals/corr/abs-1803-03467,DBLP:journals/corr/abs-1904-12575,DBLP:journals/corr/abs-1905-07854} and vary in term of size. The Knowledge Graphs (KG) for each datasets is bulit by different way. For Last.FM 2011, Book-Crossing, MovieLens-1M, KG are built by Microsoft Satori and the confidence level greater than 0.9 is set. The KG of LFM-1b 2015, Amazon-book are built by title matching which method is described in \cite{DBLP:journals/corr/abs-1807-11141}. For Yelp2018, KG is built from local business information network \cite{DBLP:journals/corr/abs-1905-07854}. The statistics of dataset are recorded on Table~\ref{tb:datasetstatistic}. We transform dataset into implicit feedback, where each entry is marked with 1 if item has been interacted or rated by user; otherwise, the entry is marked with 0. The rating threshold of MovieLens-1M is 4. For Book-Crossing and Last.FM 2011, LFM-1b 2015, Amazon-book and Yelp2018, we treat it as positive example if we observed user-item interaction. In order to ensure the quality of datasets, we apply 20-core setting for Amazon-book and Yelp 2018 and 50-core setting for LFM-1b2015. In other words, the datasets only remain users and items with at least the number of cores interactions.

\begin{itemize}
\item \textbf{MovieLens-1M} dataset is a widely used benchmark dataset in movie recommendations, the dataset contains approximately 1 million explicit ratings (ranging from 1 to 5) on total 2445 items from 6036 users. 
\item \textbf{Book-Crossing} dataset is collected from the Book-Crossing community. It contains 139746 explicit ratings (ranging from 0 to 10) on total 14967 items from 17860 users.
\item \textbf{Last.FM 2011} is the dataset about music listening collected from Last.fm online music systems. This dataset contains 42346 explicit ratings record on total 1872 items from 3846 users.
\item \textbf{LFM-1b 2015} is the dataset about music which record artists, albums, tracks, and users, as well as individual listening events information. This dataset contains about 3 million explicit ratings record on total  15471 items from 12134 users.
\item \textbf{Amazon-book} is the dataset about user's preferences on book products. It record information about user, item, rating and timestamp. This dataset contains about half million explicit ratings record on total 9854 items from 7 thousand users.
\item \textbf{Yelp2018} is the dataset from 2018 edition of Yelp challenge and is about local businesses. This dataset contains about 1.2 million explicit ratings record on total 14 thousand items from 16 thousand users.
\end{itemize}

\subsection{GNN-based Recommendation Baseline Model}

To evaluate the performance of GNN-based recommendation model with stage-wise training, we conduct experiments on two GNN-based recommendation models: RippleNet and Knowledge Graph Convolutional Networks (KGCN).

\begin{itemize}

\item \textbf{RippleNet}\cite{DBLP:journals/corr/abs-1803-03467} is a hybrid based KG-aware recommendation, which combines knowledge graph embedding regularization and path-based concept. RippleNet is a memory-network-like approach that users preference embedding is aggregated from entities embedding in KG. For each user, RippleNet would sample a fixed-size set of neighbor to predict user preference.

\item \textbf{KGCN}\cite{DBLP:journals/corr/abs-1904-12575} is also a hybrid based KG-aware recommendation model, and it has user preference embedding for each user that allow model to capture users' personalized interests from relations. In addition, we adapt label smoothness regularization on KGCN which leads to better generalization\cite{DBLP:journals/corr/abs-1905-04413}. For each user, we let the sampler of KGCN uniformly sample a fixed size set of neighbors for each entity to predict user preference.
\end{itemize}

\begin{table*}[t]
\small
\centering
\caption{The results of $AUC$ and $ACC$ score in CTR prediction on all datasets}
\scalebox{1.0}{
\begin{tabular}{ccccccccccccc}
\hline \multicolumn{1}{c}{\multirow{2}{*}{Model}} &
\multicolumn{2}{c}{MovieLens-1M} &
\multicolumn{2}{c}{Book-Crossing} &
\multicolumn{2}{c}{Last.FM 2011}&
\multicolumn{2}{c}{LFM-1b 2015}&
\multicolumn{2}{c}{Amazon-book}&
\multicolumn{2}{c}{Yelp2018} \\
\cmidrule(lr){2-3} \cmidrule(lr){4-5} \cmidrule(lr){6-7}  \cmidrule(lr){8-9} \cmidrule(lr){10-11} \cmidrule(lr){12-13} &
\multicolumn{1}{c}{\emph{AUC}} & \multicolumn{1}{c}{\emph{ACC}} &
\multicolumn{1}{c}{\emph{AUC}} & \multicolumn{1}{c}{\emph{ACC}} &
\multicolumn{1}{c}{\emph{AUC}} & \multicolumn{1}{c}{\emph{ACC}} &
\multicolumn{1}{c}{\emph{AUC}} & \multicolumn{1}{c}{\emph{ACC}} &
\multicolumn{1}{c}{\emph{AUC}} & \multicolumn{1}{c}{\emph{ACC}}
& \multicolumn{1}{c}{\emph{AUC}} & \multicolumn{1}{c}{\emph{ACC}} \\
\hline \multicolumn{1}{c}{{KGNN}} & \multicolumn{1}{c}{{.9171}} &
\multicolumn{1}{c}{{.8452}} & \multicolumn{1}{c}{{.6750}} & \multicolumn{1}{c}{{.6208}} & \multicolumn{1}{c}{{.7865}} & \multicolumn{1}{c}{{.7099}} & \multicolumn{1}{c}{{.9127}} & \multicolumn{1}{c}{{.8617}} & \multicolumn{1}{c}{{.8151}} & \multicolumn{1}{c}{{.7393}} & \multicolumn{1}{c}{{.9049}} & \multicolumn{1}{c}{{.8399}} \\
\multicolumn{1}{c}{{KGNN-SW}} & \multicolumn{1}{c}{{.9223}} &
\multicolumn{1}{c}{{.8490}} & \multicolumn{1}{c}{{.7001}} & \multicolumn{1}{c}{{.6390}} & \multicolumn{1}{c}{{.8041}} & \multicolumn{1}{c}{{.7311}} & \multicolumn{1}{c}{{.9194}} & \multicolumn{1}{c}{{.8670}} & \multicolumn{1}{c}{{.8241}} & \multicolumn{1}{c}{{.7470}} & \multicolumn{1}{c}{{.9068}} & \multicolumn{1}{c}{.8420} \\
\multicolumn{1}{c}{{RippleNet}} & \multicolumn{1}{c}{{.9276}} &
\multicolumn{1}{c}{{.8557}} & \multicolumn{1}{c}{{.7630}} & \multicolumn{1}{c}{{.6909}} & \multicolumn{1}{c}{{.8081}} & \multicolumn{1}{c}{{.7418}} & \multicolumn{1}{c}{{.9361}} & \multicolumn{1}{c}{{.8847}} & \multicolumn{1}{c}{{.8216}} & \multicolumn{1}{c}{{.7461}} & \multicolumn{1}{c}{{.9203}} & \multicolumn{1}{c}{{.8588}} \\
\multicolumn{1}{c}{{RippleNet-SW}} & \multicolumn{1}{c}{{.9423}} &
\multicolumn{1}{c}{{.8721}} & \multicolumn{1}{c}{{.7666}} & \multicolumn{1}{c}{{.6929}} & \multicolumn{1}{c}{{.8120}} & \multicolumn{1}{c}{{.7457}} & \multicolumn{1}{c}{{.9530}} & \multicolumn{1}{c}{{.9015}} & \multicolumn{1}{c}{{.9010}} & \multicolumn{1}{c}{{.8259}} & \multicolumn{1}{c}{{.9481}} & \multicolumn{1}{c}{{.8920}} \\ \hline
\end{tabular}}
\label{tb:KGNN_CTR_result}
\end{table*}

\begin{table*}[t]
\small
\centering
\caption{The results of $Recall@K$ score in top-K recommendation on all
datasets}
\scalebox{1.0}{
\begin{tabular}{p{1.8cm}|cc|cc|cc|cc|cc|cc}
\hline \multicolumn{1}{c|}{\multirow{2}{*}{Model}} &
\multicolumn{2}{c|}{MovieLens-1M} &
\multicolumn{2}{c|}{Book-Crossing} &
\multicolumn{2}{c|}{Last.FM 2011} &
\multicolumn{2}{c|}{LFM-1b 2015} &
\multicolumn{2}{c|}{Amazon-book} &
\multicolumn{2}{c}{Yelp 2018} \\
    & \multicolumn{1}{c}{\emph{R@25}} & \multicolumn{1}{c|}{\emph{R@50}} &
\multicolumn{1}{c}{\emph{R@25}} & \multicolumn{1}{c|}{\emph{R@50}} &
\multicolumn{1}{c}{\emph{R@25}} & \multicolumn{1}{c|}{\emph{R@50}} &
\multicolumn{1}{c}{\emph{R@25}} & \multicolumn{1}{c|}{\emph{R@50}} &
\multicolumn{1}{c}{\emph{R@25}} & \multicolumn{1}{c|}{\emph{R@50}} &
\multicolumn{1}{c}{\emph{R@25}} & \multicolumn{1}{c}{\emph{R@50}} \\  \hline 
\multicolumn{1}{c|}{{KGNN}} & \multicolumn{1}{c}{{.1229}} &
\multicolumn{1}{c|}{{.2102}} & \multicolumn{1}{c}{{.0483}} & \multicolumn{1}{c|}{{.0763}} & \multicolumn{1}{c}{{.1343}} & \multicolumn{1}{c|}{{.1890}} & \multicolumn{1}{c}{{.0067}} & \multicolumn{1}{c|}{{.0121}} & \multicolumn{1}{c}{{.0377}} & \multicolumn{1}{c|}{{.0603}} & \multicolumn{1}{c}{{.0395}} & \multicolumn{1}{c}{{.0634}} \\
\multicolumn{1}{c|}{{KGNN-SW}} & \multicolumn{1}{c}{{.1308}} &
\multicolumn{1}{c|}{{.2236}} & \multicolumn{1}{c}{{.0478}} & \multicolumn{1}{c|}{{.0785}} & \multicolumn{1}{c}{{.1463}} & \multicolumn{1}{c|}{{.2119}} & \multicolumn{1}{c}{{.0079}} & \multicolumn{1}{c|}{{.0134}} & \multicolumn{1}{c}{{.0405}} & \multicolumn{1}{c|}{{.0672}} & \multicolumn{1}{c}{{.0412}} & \multicolumn{1}{c}{{.0679}} \\
\multicolumn{1}{c|}{{RippleNet}} & \multicolumn{1}{c}{{.1302}} &
\multicolumn{1}{c|}{{.2300}} & \multicolumn{1}{c}{{.0482}} & \multicolumn{1}{c|}{{.0792}} & \multicolumn{1}{c}{{.1177}} & \multicolumn{1}{c|}{{.1917}} & \multicolumn{1}{c}{{.0101}} & \multicolumn{1}{c|}{{.0173}} & \multicolumn{1}{c}{{.0362}} & \multicolumn{1}{c|}{{.0618}} & \multicolumn{1}{c}{{.0400}} & \multicolumn{1}{c}{{.0692}} \\
\multicolumn{1}{c|}{{RippleNet-SW}} & \multicolumn{1}{c}{{.1339}} &
\multicolumn{1}{c|}{{.2371}} & \multicolumn{1}{c}{{.0491}} & \multicolumn{1}{c|}{{.0797}} & \multicolumn{1}{c}{{.1158}} & \multicolumn{1}{c|}{{.1917}} & \multicolumn{1}{c}{{.0123}} & \multicolumn{1}{c|}{{.0182}} & \multicolumn{1}{c}{{.0578}} & \multicolumn{1}{c|}{{.0910}} & \multicolumn{1}{c}{{.0509}} & \multicolumn{1}{c}{{.0853}} \\ 
\hline
\end{tabular}}
\label{tb:KGNN_top_k_result}
\end{table*}

\begin{table}[t]
\small
\centering
\caption{$AUC$ result of KGCN with different neighbors sampling size, where * denotes to GraphSW}
\scalebox{1.0}{
\begin{tabular}{p{2.5cm}|ccccccc}
\hline \multicolumn{1}{c|}{\multirow{1}{*}{$S$}} &  \multicolumn{1}{c}{{2}} & \multicolumn{1}{c}{{4}} &
\multicolumn{1}{c}{{8}} & \multicolumn{1}{c}{{16}} &
\multicolumn{1}{c}{{32}} & \multicolumn{1}{c}{{64}}  \\ \hline \multicolumn{1}{c|}{MovieLens-1M} & 
\multicolumn{1}{c}{{.9156}} & \multicolumn{1}{c}{{.9157}} &
\multicolumn{1}{c}{{\textbf{.9171}}} & \multicolumn{1}{c}{{.9160}} &
\multicolumn{1}{c}{{.9167}} & \multicolumn{1}{c}{{.9146}} \\ 
\multicolumn{1}{c|}{MovieLens-1M*} & 
\multicolumn{1}{c}{{.9201}} & \multicolumn{1}{c}{{.9198}} &
\multicolumn{1}{c}{{.\textbf{9223}}} & \multicolumn{1}{c}{{.9198}} &
\multicolumn{1}{c}{{.9191}} & \multicolumn{1}{c}{{.9195}} \\
\multicolumn{1}{c|}{Book-Crossing} &  \multicolumn{1}{c}{{.6689}} & \multicolumn{1}{c}{{.6694}} &
\multicolumn{1}{c}{{.6635}} & \multicolumn{1}{c}{\textbf{.6750}} &
\multicolumn{1}{c}{.6472} & \multicolumn{1}{c}{.6418} \\
\multicolumn{1}{c|}{Book-Crossing*} &  \multicolumn{1}{c}{{\textbf{.7001}}} & \multicolumn{1}{c}{{.6895}} &
\multicolumn{1}{c}{{.6893}} & \multicolumn{1}{c}{{.6771}} &
\multicolumn{1}{c}{{.6745}} & \multicolumn{1}{c}{{.6607}} \\
\multicolumn{1}{c|}{Last.FM 2011}  & 
\multicolumn{1}{c}{{.7714}} & \multicolumn{1}{c}{{.7817}} &
\multicolumn{1}{c}{{.7859}} & \multicolumn{1}{c}{{\textbf{.7865}}} &
\multicolumn{1}{c}{{.7617}} & \multicolumn{1}{c}{{.7705}} \\
\multicolumn{1}{c|}{Last.FM 2011*}  & 
\multicolumn{1}{c}{{\textbf{.8041}}} & \multicolumn{1}{c}{{.8024}} &
\multicolumn{1}{c}{{.8013}} & \multicolumn{1}{c}{{.7964}} &
\multicolumn{1}{c}{{.8002}} & \multicolumn{1}{c}{{.7989}} \\ 
\multicolumn{1}{c|}{LFM-1b 2015}  & 
\multicolumn{1}{c}{{.9117}} & \multicolumn{1}{c}{{\textbf{.9127}}} &
\multicolumn{1}{c}{{.9090}} & \multicolumn{1}{c}{{.9087}} &
\multicolumn{1}{c}{{.9086}} & \multicolumn{1}{c}{{.9085}} \\
\multicolumn{1}{c|}{LFM-1b 2015*}  & 
\multicolumn{1}{c}{{\textbf{.9194}}} & \multicolumn{1}{c}{{.9191}} &
\multicolumn{1}{c}{{.9189}} & \multicolumn{1}{c}{{.9167}} &
\multicolumn{1}{c}{{.9165}} & \multicolumn{1}{c}{{.9172}} \\
\multicolumn{1}{c|}{Amazon-book}  & 
\multicolumn{1}{c}{{\textbf{.8151}}} & \multicolumn{1}{c}{{.8122}} &
\multicolumn{1}{c}{{.8111}} & \multicolumn{1}{c}{{.8060}} &
\multicolumn{1}{c}{{.8002}} & \multicolumn{1}{c}{{.8047}} \\
\multicolumn{1}{c|}{Amazon-book*}  & 
\multicolumn{1}{c}{{\textbf{.8241}}} & \multicolumn{1}{c}{{.8224}} &
\multicolumn{1}{c}{{.8169}} & \multicolumn{1}{c}{{.8160}} &
\multicolumn{1}{c}{{.8171}} & \multicolumn{1}{c}{{.8181}} \\
\multicolumn{1}{c|}{Yelp 2018}  & 
\multicolumn{1}{c}{{.8999}} & \multicolumn{1}{c}{{.8992}} &
\multicolumn{1}{c}{{.9022}} & \multicolumn{1}{c}{{.9029}} &
\multicolumn{1}{c}{{.9024}} & \multicolumn{1}{c}{{\textbf{.9049}}} \\
\multicolumn{1}{c|}{Yelp 2018*}  & 
\multicolumn{1}{c}{{.9033}} & \multicolumn{1}{c}{{.9041}} &
\multicolumn{1}{c}{{\textbf{.9068}}} & \multicolumn{1}{c}{{.9059}} &
\multicolumn{1}{c}{{.9058}} & \multicolumn{1}{c}{{.9045}} \\\hline
\end{tabular}}
\label{tb:KGNN_auc_result_by_neighbor}
\end{table}

\begin{table}[t]
\small
\centering
\caption{$AUC$ result of RippleNet with different memory size, where * denotes to GraphSW}
\scalebox{1.0}{
\begin{tabular}{p{2.5cm}|ccccccc}
\hline \multicolumn{1}{c|}{\multirow{1}{*}{$S$}} &  \multicolumn{1}{c}{{2}} & \multicolumn{1}{c}{{4}} &
\multicolumn{1}{c}{{8}} & \multicolumn{1}{c}{{16}} &
\multicolumn{1}{c}{{32}} & \multicolumn{1}{c}{{64}}  \\ \hline \multicolumn{1}{c|}{MovieLens-1M} & 
\multicolumn{1}{c}{{.8903}} & \multicolumn{1}{c}{{.8964}} &
\multicolumn{1}{c}{{.9047}} & \multicolumn{1}{c}{{.9139}} &
\multicolumn{1}{c}{{.9214}} & \multicolumn{1}{c}{{\textbf{.9276}}} \\ 
\multicolumn{1}{c|}{MovieLens-1M*} & 
\multicolumn{1}{c}{{.8996}} & \multicolumn{1}{c}{{.9092}} &
\multicolumn{1}{c}{{.9186}} & \multicolumn{1}{c}{{.9292}} &
\multicolumn{1}{c}{{.9357}} & \multicolumn{1}{c}{{\textbf{.9423}}} \\ 
\multicolumn{1}{c|}{Book-Crossing} &  \multicolumn{1}{c}{{.7520}} & \multicolumn{1}{c}{{.7615}} &
\multicolumn{1}{c}{{\textbf{.7630}}} & \multicolumn{1}{c}{{.7602}} &
\multicolumn{1}{c}{.7531} & \multicolumn{1}{c}{.6717} \\
\multicolumn{1}{c|}{Book-Crossing*} &  \multicolumn{1}{c}{{.7596}} & \multicolumn{1}{c}{{.7647}} &
\multicolumn{1}{c}{{\textbf{.7666}}} & \multicolumn{1}{c}{{.7661}} &
\multicolumn{1}{c}{{.7513}} & \multicolumn{1}{c}{{.7412}} \\
\multicolumn{1}{c|}{Last.FM 2011}  & 
\multicolumn{1}{c}{{.7926}} & \multicolumn{1}{c}{{.7997}} &
\multicolumn{1}{c}{{.8053}} & \multicolumn{1}{c}{{\textbf{.8081}}} &
\multicolumn{1}{c}{{.8044}} & \multicolumn{1}{c}{{.8005}} \\
\multicolumn{1}{c|}{Last.FM 2011*}  & 
\multicolumn{1}{c}{{.8055}} & \multicolumn{1}{c}{{.8083}} &
\multicolumn{1}{c}{{.8112}} & \multicolumn{1}{c}{{\textbf{.8120}}} &
\multicolumn{1}{c}{{.8078}} & \multicolumn{1}{c}{{.8035}} \\
\multicolumn{1}{c|}{LFM-1b 2015}  & 
\multicolumn{1}{c}{{.8817}} & \multicolumn{1}{c}{{.8954}} &
\multicolumn{1}{c}{{.9039}} & \multicolumn{1}{c}{{.9171}} &
\multicolumn{1}{c}{{.9279}} & \multicolumn{1}{c}{{\textbf{.9361}}} \\
\multicolumn{1}{c|}{LFM-1b 2015*}  & 
\multicolumn{1}{c}{{.8902}} & \multicolumn{1}{c}{{.9065}} &
\multicolumn{1}{c}{{.9194}} & \multicolumn{1}{c}{{.9345}} &
\multicolumn{1}{c}{{.9433}} & \multicolumn{1}{c}{{\textbf{.9530}}} \\
\multicolumn{1}{c|}{Amazon-book}  & 
\multicolumn{1}{c}{{.6929}} & \multicolumn{1}{c}{{.6990}} &
\multicolumn{1}{c}{{.7334}} & \multicolumn{1}{c}{{.7706}} &
\multicolumn{1}{c}{{.7939}} & \multicolumn{1}{c}{{\textbf{.8216}}} \\
\multicolumn{1}{c|}{Amazon-book*}  & 
\multicolumn{1}{c}{{.7256}} & \multicolumn{1}{c}{{.7870}} &
\multicolumn{1}{c}{{.8404}} & \multicolumn{1}{c}{{.8693}} &
\multicolumn{1}{c}{{\textbf{.9010}}} & \multicolumn{1}{c}{{.9000}} \\
\multicolumn{1}{c|}{Yelp 2018}  & 
\multicolumn{1}{c}{{.7987}} & \multicolumn{1}{c}{{.8650}} &
\multicolumn{1}{c}{{.8955}} & \multicolumn{1}{c}{{.9064}} &
\multicolumn{1}{c}{{.9149}} & \multicolumn{1}{c}{{\textbf{.9203}}} \\
\multicolumn{1}{c|}{Yelp 2018*}  & 
\multicolumn{1}{c}{{.8649}} & \multicolumn{1}{c}{{.8987}} &
\multicolumn{1}{c}{{.9199}} & \multicolumn{1}{c}{{.9357}} &
\multicolumn{1}{c}{{.9384}} & \multicolumn{1}{c}{{\textbf{.9481}}} \\\hline
\end{tabular}}
\label{tb:RippleNet_auc_result_by_neighbor}
\end{table}

\subsection{Experiments Setup} 

We evaluate stage-wise training in two experiments settings: (1) Click-Through Rate (CTR) and (2) top-K recommendation. In CTR prediction, the trained model is applied to predict each interaction in the test set. We use AUC and ACC to evaluate CTR prediction. In top-K recommendation, trained model is applied to select K items with highest predicted click probability for each user in the test set, and Recall@K is chosen to evaluate the result. Each dataset is split into training, evaluation and test set in ratio of 6:2:2. Each experiment is repeated at least 5 time and average score is presented. For both recommendation models, all trainable parameters are optimized by Adam algorithm. 

\subsubsection{Parameter Settings}  

For KGCN and RippleNet, the hyper-parameters are determined by optimizing AUC on a validation set. For both models, the learning rate $\eta$ is selected from [0.1, 0.02, 0.005, 0.0002, 0.0005, 0.0008] and regularization weight is selected from [\(1\times {10}^{-4}\),\(1\times {10}^{-5}\),\(2\times {10}^{-5}\) \(1\times {10}^{-7}\)]. To reduce heavier computation, we set the embedding dimension of all entity and relation to 8 and 16 for RippleNet and KGCN. Moreover, in GraphSW, early stopping strategy is adpoted according to performance of current stage.

\subsection{Results}

The results of KGCN and RippleNet on CTR prediction experiment and top-K recommendation are shown on Table~\ref{tb:KGNN_CTR_result} and  Table~\ref{tb:KGNN_top_k_result}. We would discuss the recommendation performance gain from GraphSW in following:

\subsubsection{Improvement on different datasets}
In general, GraphSW improves the recommendation performance of KGCN and RippleNet on all dataset. For KGCN, the sparse datasets, such as Book-Crossing and Last.FM would cause KGCN fail to converge. However, with stage-wise training, KGCN achieve gain of AUC by 2.2$\%$ and 3.7$\%$ for Last.FM 2011 and Book-Crossing respectively. It is concluded that GraphSW assists KGCN with collecting more information on KG and performs well on sparse scenarios. In addition, for RippleNet, huge improvement of performing GraphSW is found on dataset with large size KG, such as MovieLens-1M, LFM-1b 2015, Amazon-book and Yelp2018, however, improvement is relatively small for KGCN. With GraphSW, RippleNet can collect more information on KG. As a result, the improvement from GraphSW is larger on RippleNet than on KGCN of dataset with lager KG size.

\subsubsection{Size of neighbor sampling in each hop} After comparing the improvement on different dataset. We further investigate the model's performance with GraphSW on different neighbor sampling size. The experiment result of different neighbor sampling size is shown on Table~\ref{tb:KGNN_auc_result_by_neighbor} and Table~\ref{tb:RippleNet_auc_result_by_neighbor}. For both KGCN and RippleNet, the hop number is set to 1 on all dataset. Specifically for Yelp 2018 dataset, we optimize the hop number to 2 for KGCN because of better performance. In general, GraphSW improves the model performance on different memory size. For KGCN, we surprisingly observed that KGCN achieves best result on almost every dataset with GraphSW when the neighbor sampling size is small. When KGCN is trained with small size of neighbor nodes, the model can better capture neighbor node's information in KG. Moreover, KGCN can gradually learn information from KG with GraphSW. As a result, KGCN achieves best performance  with GraphSW when the neighbor sampling size is small. For RippleNet, compared with KGCN, it achieves better result with GraphSW when neighbor sampling size is large. with the help of GraphSW, we have found that the performance on dataset with large KG size obtains huge growth.

\begin{table}[t]
\small
\centering
\caption{$AUC$ result of KGCN with different hop number $H$, where * denotes to model trained with GraphSW}
\scalebox{1.0}{
\begin{tabular}{p{2.5cm}|cccc}
\hline \multicolumn{1}{c|}{\multirow{1}{*}{$H$}} &  \multicolumn{1}{c}{{1}} & \multicolumn{1}{c}{{2}} &
\multicolumn{1}{c}{{3}} & \multicolumn{1}{c}{{4}} \\ \hline \multicolumn{1}{c|}{MovieLens-1M} & 
\multicolumn{1}{c}{{.9171}} & \multicolumn{1}{c}{{.9082}} &
\multicolumn{1}{c}{{.9037}} & \multicolumn{1}{c}{{.8915}}  \\ 
\multicolumn{1}{c|}{MovieLens-1M*} & 
\multicolumn{1}{c}{{.9223}} & \multicolumn{1}{c}{{.9164}} &
\multicolumn{1}{c}{{.9120}} & \multicolumn{1}{c}{{.9121}}  \\
\multicolumn{1}{c|}{Book-Crossing} &  \multicolumn{1}{c}{{.6750}} & \multicolumn{1}{c}{{.6644}} &
\multicolumn{1}{c}{{.6204}} & \multicolumn{1}{c}{.5628} \\
\multicolumn{1}{c|}{Book-Crossing*} &  \multicolumn{1}{c}{{.7001}} & \multicolumn{1}{c}{{.6769}} &
\multicolumn{1}{c}{{.6689}} & \multicolumn{1}{c}{{.6614}} \\
\multicolumn{1}{c|}{Last.FM 2011}  & 
\multicolumn{1}{c}{{.7865}} & \multicolumn{1}{c}{{.7608}} &
\multicolumn{1}{c}{.7213} & \multicolumn{1}{c}{{.5803}} \\
\multicolumn{1}{c|}{Last.FM 2011*}  &
\multicolumn{1}{c}{{.8041}} & \multicolumn{1}{c}{{.7853}} &
\multicolumn{1}{c}{{.7882}} & \multicolumn{1}{c}{{.7821}} \\
\multicolumn{1}{c|}{LFM-1b 2015}  & 
\multicolumn{1}{c}{{.9136}} & \multicolumn{1}{c}{{.9043}} &
\multicolumn{1}{c}{{.8998}} & \multicolumn{1}{c}{{.8975}} \\
\multicolumn{1}{c|}{LFM-1b 2015*}  & 
\multicolumn{1}{c}{{.9194}} & \multicolumn{1}{c}{{.9198}} &
\multicolumn{1}{c}{{.9183}} & \multicolumn{1}{c}{{.9128}} \\
\multicolumn{1}{c|}{Amazon-book}  & 
\multicolumn{1}{c}{{.8151}} & \multicolumn{1}{c}{{.8042}} &
\multicolumn{1}{c}{{.7907}} & \multicolumn{1}{c}{{.7622}}  \\
\multicolumn{1}{c|}{Amazon-book*}  & 
\multicolumn{1}{c}{{.8241}} & \multicolumn{1}{c}{{.8232}} &
\multicolumn{1}{c}{{.8138}} & \multicolumn{1}{c}{{.8245}} \\
\multicolumn{1}{c|}{Yelp 2018}  & 
\multicolumn{1}{c}{{.9009}} & \multicolumn{1}{c}{{.8782}} &
\multicolumn{1}{c}{{.8720}} & \multicolumn{1}{c}{{.7609}}\\
\multicolumn{1}{c|}{Yelp 2018*}  & 
\multicolumn{1}{c}{{.9068}} & \multicolumn{1}{c}{{.9066}} &
\multicolumn{1}{c}{{.9038}} & \multicolumn{1}{c}{{.9051}}  \\\hline
\end{tabular}}
\label{tb:KGNN_top_k_result_by_hop}
\end{table}

\begin{table}[t]
\small
\centering
\caption{$AUC$ result of KGCN as the hop number $H$ is 4, where * denotes to GraphSW with transferring whole training parameters}
\scalebox{1.0}{
\begin{tabular}{p{2.0cm}ccc}
\hline \multicolumn{1}{c}{\multirow{1}{*}{Model}} &
\multicolumn{1}{c}{MovieLens-1M} &
\multicolumn{1}{c}{Book-Crossing} &
\multicolumn{1}{c}{Last.FM 2011}
\\ \hline
\multicolumn{1}{c}{{KGNN-SW*}} & \multicolumn{1}{c}{{.8868}} &
\multicolumn{1}{c}{{.6476}} & \multicolumn{1}{c}{{.7372}} \\
\multicolumn{1}{c}{{KGNN-SW}} & \multicolumn{1}{c}{{.9121}} &
\multicolumn{1}{c}{{.6614}} & \multicolumn{1}{c}{{.7821}} \\ \hline
\multicolumn{1}{c}{\multirow{1}{*}{Model}} & \multicolumn{1}{c}{LFM-1b 2015}& \multicolumn{1}{c}{Amazon-book}& 
\multicolumn{1}{c}{Yelp 2018} \\ \hline 
\multicolumn{1}{c}{{KGNN-SW*}} & \multicolumn{1}{c}{{.9070}} &
\multicolumn{1}{c}{{.7780}} & \multicolumn{1}{c}{{.8911}} \\
\multicolumn{1}{c}{{KGNN-SW}} & \multicolumn{1}{c}{{.9194}} &
\multicolumn{1}{c}{{.8245}} & \multicolumn{1}{c}{{.9062}} \\ \hline
\end{tabular}}
\label{tb:KGNN_tr_whole_para_result}
\end{table}
 
\subsubsection{Improvement on different hop number} As we mentioned before, KGCN fails to converge as the hop number increases \cite{DBLP:journals/corr/abs-1904-12575} due to increasing noise and the growth of parameter dimensionality. As a result, we conduct experiment to investigate improvement from GraphSW on different hop number. KGCN's result is shown in Table~\ref{tb:KGNN_top_k_result_by_hop} and RippleNet's result is shown in Appendix. In general, GraphSW improves both models' performance on different hop number. Moreover, we found that KGCN would easily collapse when hop number is 3 or 4, but RippleNet seems not. However, with GraphSW, KGCN's performance would be improved by a large margin. The average improvement on hop number 4 is 34.8$\%$, 17.5$\%$, 2.3$\%$, 2.2$\%$, 8.2$\%$ and 0.9$\%$ for Last.Fm 2011, Book-Crossing, MovieLens-1M, LFM-1b 2015, Amazon-book and Yelp 2018 respectively. To further investigate GraphSW on KGCN when hop number is large, we conduct another experiment to transfer whole training parameters instead of only transferring KG representation in different stages and the result is showed on Table~\ref{tb:KGNN_tr_whole_para_result}, and the performance drops significantly when transferring whole model parameters. In conclusion, KG representation with comprehensive view of graph can benefit KGCN training process when training KGCN’s parameter in different stages.
\section{Conclusion}
In this paper, we proposed GraphSW, a training protocol based on stage-wise training for GNN-based recommendation model. With GraphSW, we conduct comprehensive study on performance gain when different number of KG information is used. In general, we find GraphSW improves KGCN and RippleNet on every dataset and it shows that the resampling strategy should not be neglected by GNN-based recommendation model which would sample fixed-size set of neighbors. In addition, it is found that KGCN and RippleNet achieve best result on different situation because of the different aggregating method. For KGCN, which has structure similar to PinSage, recommendation's performance achieves best result on almost every dataset with stage-wise training when the neighbor sampling size is small. In addition, stage-wise training allows us to train model in different stage. In the absence of stage-wise training, KGCN suffers from noise and fails to converge as hop number increases because of the growth of parameter dimensionality. However, with stage-wise training, the performance of KGCN is improved by a large margin. The improvement brought from GraphSW instructs us to design GNN-based recommendation model with more complicate architecture and parameters.

\nocite{*} 
\bibliographystyle{unsrt}
\bibliography{main} 


\end{document}